\documentclass[%
 reprint,
 amsmath,amssymb,
 aps,superscriptaddress
]{revtex4-2}

\usepackage{graphicx}
\usepackage{dcolumn}
\usepackage{bm}
\usepackage[normalem]{ulem}
\usepackage{xcolor}
\usepackage{mathrsfs}
\usepackage[T1]{fontenc}
\usepackage{inputenc}
\usepackage[version=4]{mhchem}
\usepackage[slantedGreek]{mathpazo}
\usepackage{hyperref}
\usepackage{float}
\usepackage{soul}
\begin{document}
\newcommand{\rb}[1]{\textcolor{red}{\it#1}}
\newcommand{\rbout}[1]{\textcolor{red}{\sout{#1}}}

\preprint{APS/123-QED}

\title{Relativistic KRCI calculations of symmetry violating interaction constants for YbX (X: Cu, Ag and Au) molecules}

\author{Ankush Thakur}
\email{ankush\_t@ph.iitr.ac.in}
\altaffiliation{Contributed equally to the work}
\affiliation{Department of Physics, Indian Institute of Technology Roorkee, Roorkee-247667, India}
\author{Renu Bala}
\email{ renub@umk.pl}
\altaffiliation{Contributed equally to the work}
\affiliation{Institute of Physics, Faculty of Physics, Astronomy and Informatics, Nicolaus Copernicus University, Grudziądzka 5, 87-100 Toru\'n, Poland}
\author{H. S. Nataraj}
\affiliation{Department of Physics, Indian Institute of Technology Roorkee, Roorkee-247667, India}

\begin{abstract}
The present work reports the parity ($\mathcal{P}$)-odd and time-reversal ($\mathcal{T}$)-odd interaction constants for the ground electronic state, X\,$^2\Sigma^{+}_{1/2}$, of YbX, X: Cu, Ag and Au molecules. The reported results have been calculated using the Kramers-restricted configuration interaction method limited to single and double excitations, in conjunction with relativistic core-valence double-, triple-, and quadruple-zeta quality basis sets, within a four-component relativistic framework. The computed results for the symmetry violating properties have been compared with the available results in the literature. Further, the parallel and perpendicular components of the hyperfine structure constants for the constituent atoms in YbX molecules are reported here for the first time.

\begin{description}
\item[Keywords]
parity and time-reversal odd interaction constants, effective electric field, electric dipole moment of an electron, hyperfine structure constants, Kramers-restricted configuration interaction.
\end{description}
\end{abstract}

\maketitle

\section{\label{sec:section1}Introduction}

The study of fundamental symmetry violations has significant implications in probing the physics beyond the Standard Model (BSM) of elementary particles~\cite{Ginges_2004,Fukuyama_2012,Safronova_2018}. An intrinsic electric dipole moment of the electron (eEDM, denoted by $d_e$) is a direct manifestation of the simultaneous violation of parity ($\mathcal{P}$) and time-reversal ($\mathcal{T}$) symmetries~\cite{Purcell_1950,Landau_1957}. Violation of $\mathcal{T}$ symmetry implies that the non-zero eEDM involves $\mathcal{CP}$ symmetry violation ($\mathcal{C}$ refers to charge-conjugation symmetry) on the basis of the $\mathcal{CPT}$ theorem~\cite{Luders_2000}. This is of considerable interest for testing and constraining new theories proposed to explain the observed imbalance between matter and antimatter (\textit{viz.} baryon asymmetry) in the Universe~\cite{Kazarian_1992,Virdee_2016,Chupp_2019}. \\

An additional source contributing to $\mathcal{P}$- \& $\mathcal{T}$- symmetry violation arises from the scalar–pseudoscalar (S–PS) nucleon–electron neutral current interaction~\cite{Quiney_1998}. The strength of this interaction is characterized by the S-PS interaction constant, $k_s$, and its estimation is of great importance for understanding the BSM physics, analogous to $d_e$. \\

Over the past few decades, there has been remarkable progress in both experimental and theoretical efforts in the search for the eEDM. In this context, heavy open-shell polar molecules are preferred over atoms owing to their large effective electric fields ($\varepsilon_{eff}$)~\cite{Sandars_1967,Prasannaa_2015}. Many ongoing experiments employing diatomic molecules such as ThO$^*$~\cite{Andreev_2018}, HfF$^+$~\cite{Roussy_2023}, YbF~\cite{Fitch_2020}, and BaF~\cite{Aggarwal_2018} have yielded the most stringent constraints on the eEDM to date. The interpretation of experimental results requires precise knowledge of $\mathcal{P}$-odd \& $\mathcal{T}$-odd interaction constants. These constants include $W_d$, which is related to $d_e$ via $\varepsilon_{eff}$, and $W_s$, which characterizes the nucleon-electron S-PS interaction. The accurate evaluation of these constants relies entirely on calculations based on relativistic many-body theories.\\

Among the wide range of molecular candidates proposed for eEDM searches~\cite{Kudashov_2014,Prasannaa_2015,Fleig_2016,Fazil_2019,Ramanuj_2021,Fleig_2021}, YbX (X = Cu, Ag and Au) molecules have been the subject of recent investigations. To mention, Verma \textit{et al.}~\cite{Verma_2020} identified YbAg as a promising candidate for eEDM experiments using clock transitions. Owing to this, Yuan and Liu~\cite{Yuan_2024} have examined the formation of ultracold YbAg molecules via photoassociation using \textit{ab initio} calculations. Quite recently, $W_d$ and $W_s$ constants for YbX molecules have been reported by Polet \textit{et al.}~\cite{Polet_2024} using the coupled-cluster approach. To the best of our knowledge, this is the only work that reports the symmetry violating constants of these molecules. However, calculations of potential energy curves, permanent electric dipole and quadrupole moments, and static electric dipole polarizabilities for YbX molecules in their ground electronic states have been reported in Ref.~\cite{Tomza_2021}.\\ 

In the current work, we have carried out relativistic calculations of $\mathcal{P}$-odd \& $\mathcal{T}$-odd interaction constants of YbX molecules by utilizing the Kramers-restricted configuration interaction method limited to single and double excitations (KRCISD) together with the relativistic basis sets. The theoretical calculation of hyperfine structure (HFS) constants, similar to the eEDM, requires an accurate wavefunction in the near-nuclear region. Therefore, we have also computed the parallel and perpendicular components of the magnetic dipole HFS constants for different isotopes of the atoms constituting the YbX molecules.\\

The paper is organized into four subsequent sections: the introduction is followed by a detailed description of the theory in Section~\ref{section2}, computational details in Section~\ref{section3}, a brief discussion of the results in Section~\ref{section4}, and finally, the summary of the current work in Section~\ref{section5}.

\section{\label{section2}Theory}

\subsection{$\mathcal{P}$-odd \& $\mathcal{T}$-odd interaction constant relevant to eEDM}

In a molecular system, the internal electric field experienced by an electron due to the internal charge distribution of electrons and nuclei can be termed as an effective electric field ($\varepsilon_{eff}$). The expectation value of the operator representing the interaction of electron electric dipole moment (eEDM) with $\varepsilon_{eff}$ in a molecular system is expressed as~\cite{Fleig_2014,Prasannaa_2015,Sunaga_2019,Bala_2023},
\begin{eqnarray}
 \Delta \textit{U} = \bigg\langle\sum\limits_{j = 1}^{N_e}H_{EDM}(j)\bigg\rangle_{\Psi}&=&-d_e\bigg\langle\sum\limits_{j = 1}^{N_e} \gamma_j^0\,\vec{\Sigma}_j \cdot \vec{\varepsilon}_j\bigg\rangle_{\Psi}, \nonumber\\
&=&-\frac{2icd_e}{e\hbar}\bigg\langle\sum\limits_{j = 1}^{N_e}\gamma_j^0\,\gamma_j^5\,\vec{p}_j^2\bigg\rangle_{\Psi},
\end{eqnarray}
where $N_e$ is the number of electrons; $d_e$ is the intrinsic electric dipole moment of an electron; $\vec{\Sigma}_j$ denotes the four-component Pauli spin matrices; $\gamma^5=i\gamma^0\gamma^1\gamma^2\gamma^3$, with $\gamma^0, \gamma^1, \gamma^2$, and $\gamma^3$, representing the four-component Dirac matrices; $\vec{\varepsilon}_j$ is the electric field at the position of $j^{th}$ electron; $\vec{p}_j$ is the momentum operator; $c$ represents the speed of light; $\hbar$ is the Planck constant $h$ divided by $2\pi$; $e$ is the charge of the electron, and $\Psi$ is the relativistic wavefunction of the ground state of YbX molecules, obtained from the many-body theory. Finally, the $\varepsilon_{eff}$ experienced by the unpaired electron in the molecular system is given by~\cite{Fleig_2013},
\begin{eqnarray}
 \varepsilon_{eff}\,=\,W_d\,\Omega\,, 
\end{eqnarray}
where $W_d\,=\,({2ic}/{\Omega\,e\hbar})\langle\gamma^0\,\gamma^5\,p^2\rangle_{\Psi}$ is the $\mathcal{P}$-odd \& $\mathcal{T}$-odd interaction constant and $\Omega$ is the 
component of the total angular momentum for the ground state of a given 
molecule along the $z$-axis of the coordinate system. For the molecular systems considered in this work, the value of $\Omega$ is 1/2. The intrinsic eEDM value is determined from the experimentally measured energy shift ($\Delta$\textit{U}) of the electronic state of a molecule together with the theoretically calculated $\varepsilon_{eff}$, through the relation $\Delta\textit{U}$\,=\,$-$$d_e$\,$\varepsilon_{eff}$~\cite{Abe_2014}.

\subsection{Scalar-Pseudoscalar interaction constant}

Another $\mathcal{P}$-odd \& $\mathcal{T}$-odd interaction constant, $W_s$, arising from the nucleon-electron scalar-pseudoscalar (S-PS) interaction is defined as~\cite{Skripnikov_2013},
\begin{eqnarray}
W_s &=& \frac{1}{k_{s,A}\,\Omega} \left\langle H_{\mathrm{\text{S-PS}}} \right\rangle_{\Psi}.
\end{eqnarray}
Here, $k_{s,A}$ is a dimensionless electron-nucleus S-PS coupling constant of an atom $A$, defined as~\cite{Denis_2015},
\begin{eqnarray}
k_{s,A} &=& k_{s,p} + \frac{N_A}{Z_A}\,k_{s,n}\,,
\end{eqnarray}
where $Z_A$ and $N_A$ denote the number of protons and neutrons, respectively. $k_{s,p}$ and $k_{s,n}$ are the S-PS coupling constants for electron-proton, and electron-neutron interactions, respectively.\\

The expression for the S-PS interaction Hamiltonian for a molecular system is given by~\cite{Nayak_2007,Bala_2020,Chamorro_2024},
\begin{eqnarray}
H_{\text{S-PS}} &=& \frac{i}{e}\frac{G_F}{\sqrt{2}} \sum_{j=1}^{N_e} \sum_{A=1}^{N_N} k_{s,A}\, Z_A \,\gamma^0\,\gamma^5\,\rho_A(\vec{r}_{Aj}),
\end{eqnarray}
where $\rho_A$ is the nuclear charge density normalized to unity; $G_F$ ($ = 2.22249 \times 10^{-14}\,E_h\,a_0^3$) is the Fermi coupling constant; $N_N$ represents the total number of nuclei; $\vec{r}_{Aj}$ is the distance between $A^{th}$ nucleus and $j^{th}$ electron and the summation indices $j$ and $A$ span over the number of electrons and nuclei, respectively.

\subsection{Magnetic dipole hyperfine structure constants}

The magnetic hyperfine structure (HFS) in atomic and molecular systems arises from the interaction of the nuclear magnetic dipole moment with the internally generated magnetic field of the electrons~\cite{Lindgren_2012}. The magnetic vector potential $\vec{\mathcal{A}} (\vec{r})$ at a distance $\vec{r}$ due to a nucleus of an atom is given by~\cite{Fleig_2014},
\begin{eqnarray}
\vec{\mathcal{A}} (\vec{r}) &=& \frac{\mu_0}{4\pi} \frac{\vec{\mu} \times \vec{r}}{r^3},
\end{eqnarray}
where $\mu_0$ is the vacuum permeability and $\vec{\mu}$ is the magnetic moment of nucleus.\\

The HFS Hamiltonian of an atom due to $\vec{\mathcal{A}} (\vec{r})$ is defined in the Dirac theory as~\cite{Sasmal_2015},
\begin{eqnarray}
H_{\text{HFS}} &=& -e\,c \sum_{j=1}^{N_e} \vec{\alpha}_j \cdot \vec{\mathcal{A}}_j(\vec{r}),
\end{eqnarray}
where $\vec{\alpha}_j$ denotes the Dirac matrices for $j^{th}$ electron.\\

By considering the internuclear axis of the molecular systems to be aligned along the $z$-axis, the expectation values of $z$ and $x$ (or $y$) projections of the HFS Hamiltonian yield the parallel ($A_\parallel$) and perpendicular ($A_\perp$) components of the HFS constants as~\cite{Sasmal_2016,Tulukdar_2019},
\begin{eqnarray}
A_{\parallel} &=& \frac{1}{\mathcal{I}\,\Omega}
\left\langle \Psi_{\Omega} \left| H_{\text{HFS}} \right| \Psi_{\Omega} \right\rangle, \nonumber
\\
&=& \frac{\mu_0\,e\,c}{4\pi\,\mathcal{I}\,\Omega}\,\vec{\mu} \cdot
\left\langle \Psi_{\Omega} \left|
\sum_{j=1}^{N_e}
\left( \frac{\vec{\alpha}_j \times \vec{r}_j}{r_j^3} \right)_z
\right| \Psi_{\Omega} \right\rangle, 
\end{eqnarray}
and
\begin{eqnarray}
A_{\perp} &=& \frac{1}{\mathcal{I}\,\Omega}
\left\langle \Psi_{\Omega} \left| H_{\text{HFS}} \right| \Psi_{-\Omega} \right\rangle, \nonumber
\\
&=&\frac{\mu_0\,e\,c}{4\pi\,\mathcal{I}\,\Omega}\,\vec{\mu} \cdot
\left\langle \Psi_{\Omega} \left|
\sum_{j=1}^{N_e}
\left( \frac{\vec{\alpha}_j \times \vec{r}_j}{r_j^3} \right)_{(x/y)}
\right| \Psi_{-\Omega} \right\rangle,
\end{eqnarray}
respectively. Here, $\mathcal{I}$ is the nuclear spin quantum number, and $\Psi_{\Omega}$ is the wavefunction for the ground electronic state of YbX molecules, \textit{viz.}, $^2\Sigma$ state with $\Omega\,=\,1/2$. The $^2\Sigma$ molecular electronic states with $\Omega\,=\,+1/2$ and $-1/2$ are degenerate, and their corresponding determinants differ only by the spin of an electron.

\subsection{Kramers-restricted configuration interaction method}

The four-component many-body wavefunction employed to compute the molecular interaction constants discussed above is calculated using the relativistic KRCI method. The CI wavefunction is expressed as a linear combination of determinantal functions as~\cite{Szabo},
\begin{eqnarray}
\arrowvert \Psi_{\text{CI}}\rangle &=& C_0 \arrowvert \Phi_0 \rangle + \sum_{a,p}C_a^pa_p^\dag a_a\arrowvert \Phi_0\rangle\\ \nonumber
&+&\sum_{ab,pq}C_{ab}^{pq}a_p^{\dag}a_q^{\dag}a_ba_a\arrowvert \Phi_0\rangle + \dots\,, 
\end{eqnarray}
where $\arrowvert \Phi_0 \rangle$ is the reference Dirac-Hartree-Fock wave function. The subscripts $a$, $b$, $\ldots$ denote the filled spin-orbitals, and $p$, $q$, $\ldots$ represent the virtual spin-orbitals. Thus, the operator $a_p^\dag a_a$ represents the simultaneous annihilation of an electron from the occupied spin-orbital $a$, accompanied by the creation of an electron in the virtual spin-orbital $p$. Consequently, the resulting determinant is a singly excited determinant, $|\Phi_a^p\rangle$ ($\equiv a_p^\dag a_a|\Phi_0\rangle$), and the corresponding excitation amplitude is $C_a^p$. Similarly, the third term involves a doubly excited determinant, $|\Phi_{ab}^{pq}\rangle$, with the associated excitation amplitude $C_{ab}^{pq}$.\\

In relativistic calculations, the Kramer-restricted wavefunction can be expanded in terms of a $P$-string of $j$ Kramers four-component spinors ${\{\Phi_a\}}$ and a $\bar{Q}$-string comprising of ($N-j$) Kramers time-reversal partners ${\{\Phi_{\bar{a}}\}}$, respectively as~\cite{Knecht_2010,Fleig_2016},
\begin{eqnarray}
|\Psi_K\rangle\,=\,\sum_{I=1}^{dimF(M,N)}C_{KI}\,|(P\bar Q)_I\rangle, 
\end{eqnarray}
where $dimF(M, N)$ is the dimension of the truncated $N$-particle Fock-space sector over $M$ molecular four-spinors and $C_{KI}$ are the expansion coefficients. The Kramer partners $\{\Phi_a, \Phi_{\bar{a}}\}$ are interrelated through the action of the time-reversal operator $\hat{K}$, satisfying $\hat{K} \Phi_a = \Phi_{\bar{a}}$ and $\hat{K}{\Phi_{\bar{a}}} = -\Phi_a$. The determinants $|(P\bar Q)_I\rangle$ can be expressed by strings of creation operators in second quantized form as,
\begin{eqnarray}
|(P\bar Q)\rangle=P^{\dag}\bar Q^{\dag}|\Phi_0\rangle, 
\end{eqnarray}
where $P^\dag \arrowvert \Phi_0\rangle=a_{P_1}^\dag a_{P_2}^\dag \dots a_{P_j}^\dag \arrowvert \Phi_0\rangle$ and $\bar Q^\dag \arrowvert \Phi_0 \rangle = a_{\bar Q_1}^\dag {a_{\bar{Q_2}}}^\dag \dots a_{\bar{Q}_{N-j}}^\dag \arrowvert \Phi_0\rangle$.

\section{\label{section3}Computational Details}

All molecular calculations have been performed using the KRCI module available in the DIRAC software suite~\cite{DIRAC}. In order to carry out KRCISD calculations, the generalized active space (GAS) technique~\cite{TFleig_2006} is employed to efficiently account for electron correlation effects. The Gaussian charge distribution for the nuclei is used in this work. Further, we have utilized uncontracted Dyall's core-valence double-zeta (cv2z), triple-zeta (cv3z), and quadruple-zeta (cv4z) basis sets~\cite{Dyall_2007,Gomes_2010,Dyall_2010,Dyall_2023}. The details of the basis functions used for the constituent atoms of the diatomic molecules examined in this study are given in Table~\ref{table-I}. The values of equilibrium bond lengths used in this work are~\cite{Polet_2024}: \(2.7543\) {\AA} for YbCu, \(2.8589\) {\AA} for YbAg, and \(2.6524\) {\AA} for YbAu.\\

\begin{table}[ht]
    \centering
    \caption{\label{table-I}Details of the basis functions.}
    \begin{tabular}{c c c c c}
    \hline\hline
       Atom & & Basis & & Basis functions \\
       \hline
       \multicolumn{4}{c}{\vspace{-3mm}}\\
       Yb && cv2z && 24s, 19p, 13d, 8f, 2g \\
       && cv3z && 30s, 24p, 16d, 11f, 3g, 2h \\
       && cv4z && 35s, 30p, 19d, 13f, 5g, 4h, 2i \\
       \hline
       \multicolumn{4}{c}{\vspace{-3mm}}\\
       Cu && cv2z && 15s, 11p, 6d, 2f \\
       && cv3z && 23s, 16p, 9d, 4f, 2g \\
       && cv4z && 30s, 20p, 12d, 6f, 4g, 2h \\
       \hline
       \multicolumn{4}{c}{\vspace{-3mm}}\\
       Ag && cv2z && 21s, 14p, 10d, 3f \\
       && cv3z && 28s, 20p, 13d, 5f, 3g \\
       && cv4z && 33s, 25p, 17d, 7f, 5g, 3h \\
       \hline
       \multicolumn{4}{c}{\vspace{-3mm}}\\
       Au && cv2z && 24s, 19p, 12d, 9f, 1g \\
       && cv3z && 30s, 24p, 15d, 11f, 4g, 1h \\
       && cv4z && 34s, 30p, 19d, 13f, 7g, 4h, 1i \\
       \hline\hline
    \end{tabular}
    \label{tab:my_label}
\end{table}

The molecular orbitals with energies lower than \(-1\) $E_h$ are treated as frozen core which results in \(27\) active electrons for all basis sets. To facilitate comparison with the available literature~\cite{Polet_2024}, a cutoff of \(-2\) $E_h$ is considered for the cv2z basis set alone, which increases the number of active electrons to 33. Within the GAS technique, we have considered two distinct configurations. In GAS configuration I (G-C1), the active orbital space is partitioned into three subspaces: paired (GAS1), unpaired (GAS2), and virtual orbitals (GAS3). The GAS2 subspace is occupied by an unpaired electron of the X atoms. In contrast, the second GAS configuration (G-C2) involves a redistribution of active orbitals between GAS1 and GAS2, in which GAS2 includes 6s$^2$ electrons of Yb along with an unpaired electron of the X atoms, similar to the scheme used in Ref.~\cite{Yuan_2024} for the YbAg molecule. The number of determinants along with the number of active orbitals in each GAS for the two different configurations are shown in Table~\ref{table-II}. For G-C2, the number of determinants reduces drastically in comparison to G-C1. It can also be seen that the number of determinants increases significantly even with a small increase in the size of the configuration space both in G-C1 and G-C2. Consequently, we could not enlarge the configuration space further, beyond what is considered, for correlation calculations due to the limited availability of computational resources.\\

\begin{table*}[]
\caption{\label{table-II}
Generalized active space (GAS) model for the CI wavefunctions of YbX molecules with different basis sets. The energy cutoff for the active space is fixed at \(-2\) $E_h$ for the cv2z basis set, while \(-1\) $E_h$ for the cv3z and cv4z basis sets.} 
\begin{ruledtabular}
\begin{tabular}{ccccccccccc}
&&& \multicolumn{4}{c}{GAS configuration I (G-C1)} & \multicolumn{4}{c}{GAS configuration II (G-C2)} \\
\cline{4-7}\cline{8-11}
\multicolumn{4}{c}{\vspace{-3mm}}\\
Molecule & Basis set & Virtual cutoff & GAS1 & GAS2 & GAS3 & Number of & GAS1 & GAS2 & GAS3 & Number of \\
&&&&&& determinants &&&& determinants \\
\hline
\multicolumn{4}{c}{\vspace{-3mm}}\\
\textbf{YbCu} & cv2z & 2 $E_h$ & 16 & 1 & 58 & 823281 & 15 & 2 & 58 & 166409 \\\\

     & cv3z & 1 $E_h$ & 13 & 1 & 55 & 485651 & 12 & 2 & 55 & 121078 \\
     & & 2 $E_h$ & 13 & 1 & 80 & 975541 & 12 & 2 & 80 & 239951 \\\\

     & cv4z & 1 $E_h$ & 13 & 1 & 69 & 735750 & 12 & 2 & 69 & 181711 \\
     & & 1.5 $E_h$ & 13 & 1 & 95 & 1318456 & 12 & 2 & 95 & 321660 \\
\hline
\multicolumn{4}{c}{\vspace{-3mm}}\\
\textbf{YbAg} & cv2z & 2 $E_h$ & 16 & 1 & 58 & 823281 & 15 & 2 & 58 & 166409 \\\\

     & cv3z & 1 $E_h$ & 13 & 1 & 59 & 543716 & 12 & 2 & 59 & 134532 \\
     & & 2 $E_h$ & 13 & 1 & 81 & 1002945 & 12 & 2 & 81 & 246876 \\\\

     & cv4z & 1 $E_h$ & 13 & 1 & 80 & 975541 & 12 & 2 & 80 & 239951 \\
     & & 1.5 $E_h$ & 13 & 1 & 101 & 1465373 & 12 & 2 & 101 & 356055 \\
\hline
\multicolumn{4}{c}{\vspace{-3mm}}\\
\textbf{YbAu} & cv2z & 2 $E_h$ & 16 & 1 & 58 & 823281 & 15 & 2 & 58 & 166409 \\\\

              & cv3z & 1 $E_h$ & 13 & 1 & 58 & 523558 & 12 & 2 & 58 & 129416 \\
              & & 2 $E_h$ & 13 & 1 & 89 & 1152529 & 12 & 2 & 89 & 280233 \\\\

              & cv4z & 1 $E_h$ & 13 & 1 & 78 & 866605 & 12 & 2 & 78 & 216048 \\
              & & 1.5 $E_h$ & 13 & 1 & 105 & 1477239 & 12 & 2 & 105 & 362704 \\
\end{tabular}
\end{ruledtabular}
\end{table*}

The values of nuclear spin quantum number ($\mathcal{I}$) and nuclear magnetic moment ($\vec{\mu}$) for the atoms constituting the considered molecules and their different isotopes used to calculate the HFS constants are tabulated in Table~\ref{table-III}~\cite{isotopes}. 

\begin{table}[ht]
    \centering
    \caption{\label{table-III}Nuclear spin quantum number ($\mathcal{I}$) and nuclear magnetic moment ($\vec{\mu}$) for different isotopes of Yb, Cu, Ag, and Au atoms~\cite{isotopes}.}
    \begin{tabular}{c c c c c c c}
    \hline\hline
    \multicolumn{4}{c}{\vspace{-3mm}}\\
       Atom & & & $\mathcal{I}$ &&& $\vec{\mu}/\mu_N$ \\
       \hline
       \multicolumn{4}{c}{\vspace{-3mm}}\\
       $^{171}$Yb &&& 1/2 &&& 0.4919 \\
       $^{173}$Yb &&& 5/2 &&& -0.6776 \\
       $^{63}$Cu &&& 3/2 &&& 2.2233 \\
       $^{65}$Cu &&& 3/2 &&& 2.3817 \\
       $^{107}$Ag &&& 1/2 &&& -0.11357 \\
       $^{109}$Ag &&& 1/2 &&& -0.1306905 \\
       $^{197}$Au &&& 3/2 &&& 0.148159 \\
       \hline\hline
    \end{tabular}
    \label{tab:my_label}
\end{table}

\section{\label{section4}Results and discussion}
\subsection{$\mathcal{P}$-odd and $\mathcal{T}$-odd interaction constants}
\begin{table}[ht]
    \centering
    \caption{\label{table-IV}$\mathcal{P}$-odd \& $\mathcal{T}$-odd interaction constants ($W_{d} \, (\text{in}\,\, 10^{24} h\,\text{Hz}/e\,\text{cm})$ and $W_\mathrm{s} \, (\text{in}\,\, h\,\text{kHz})$) for the ground state of YbX molecules calculated using v2z basis set within a configuration space of $\pm$ 2$E_h$.}
    \begin{tabular}{cccccccccc}
    \hline\hline
    \multicolumn{4}{c}{\vspace{-3mm}}\\
    && \multicolumn{2}{c}{G-C1} && \multicolumn{2}{c}{G-C2} && \multicolumn{2}{c}Literature~\cite{Polet_2024} \\
     \cline{3-4}\cline{6-7}\cline{9-10}
    \multicolumn{4}{c}{\vspace{-3mm}}\\
       Molecule && $|W_d|$ & $|W_s|$ && $|W_d|$ & $|W_s|$ && $|W_d|$ & $|W_s|$ \\
       \hline
       \multicolumn{4}{c}{\vspace{-3mm}}\\
       \textbf{YbCu} && 11.009 & 38.059 && 11.686 & 40.385 && 11.323 & 39.162 \\
       \textbf{YbAg} && 10.725 & 37.397 && 11.086 & 38.912 && 10.415 & 36.897 \\
       \textbf{YbAu} && 6.888 & 19.343 && 4.557 & 8.051 && 1.072 & 7.385 \\
       \hline\hline
    \end{tabular}
    \label{tab:my_label}
\end{table}

\begin{table*}
\caption{\label{table-V} $\mathcal{P}$-odd and $\mathcal{T}$-odd interaction constants ($W_{d} \, (\text{in}\,\, 10^{24} h\,\text{Hz}/e\,\text{cm})$ and $W_\mathrm{s} \, (\text{in}\,\, h\,\text{kHz})$) for the ground state of YbX molecules, calculated at the KRCISD level of theory using larger basis sets. The results presented in bold fonts are our recommended values.}
\begin{ruledtabular}
\begin{tabular}{cccccccc}
&&&& \multicolumn{2}{c}{G-C1} & \multicolumn{2}{c}{G-C2} \\
\cline{5-6}\cline{7-8}
\multicolumn{6}{c}{\vspace{-3mm}}\\
Molecule & Basis & Active electrons & Virtual orbitals & $|W_d|$ & $|W_s|$ & $|W_d|$ & $|W_s|$ \\
\hline
\multicolumn{6}{c}{\vspace{-3mm}}\\
\textbf{YbCu} & \textbf{cv3z} & 27 (-1 $E_h$) & 55 (1 $E_h$) & 10.574 & 38.411 &                   10.488 & 38.130 \\
              & & 27 & 80 (2 $E_h$) & 10.745 & 38.989 & \textbf{10.837} & \textbf{39.396} \\\\
              
              & cv4z & 27 & 69 (1 $E_h$) & 10.543 & 38.468 & 10.433 & 38.079 \\
              & & 27 & 95 (1.5 $E_h$) & 10.709 & 39.057 & 10.720 & 39.149 \\
\hline
\multicolumn{6}{c}{\vspace{-3mm}}\\
\textbf{YbAg} & \textbf{cv3z} & 27 & 59 (1 $E_h$) & 10.032 & 37.044 & 9.825 & 36.487 \\ 
              & & 27 & 81 (2 $E_h$) & 10.323 & 37.983 & \textbf{10.134} & \textbf{37.599} \\\\
              
              & cv4z & 27 & 80 (1 $E_h$) & 9.949 & 36.900 & 9.721 & 36.249 \\ 
              & & 27 & 101 (1.5 $E_h$) & 10.174 & 37.679 & 9.976 & 37.155 \\
\hline 
\multicolumn{6}{c}{\vspace{-3mm}}\\
\textbf{YbAu} & \textbf{cv3z} & 27 & 58 (1 $E_h$) & 4.438 & 9.988 & 2.878 & 2.785 \\
              & & 27 & 89 (2 $E_h$) & 5.602 & 15.222 & \textbf{3.364} & \textbf{4.830} \\\\
              
              & cv4z & 27 & 78 (1 $E_h$) & 3.423 & 25.395 & 3.414 & 25.359 \\
              & & 27 & 105 (1.5 $E_h$) & 4.032 & 28.307 & 4.024 & 28.275 \\
\end{tabular}
\end{ruledtabular}
\end{table*}

The computed results for $\mathcal{P}$-odd \& $\mathcal{T}$-odd interaction constants, $W_d$ and $W_s$, at KRCISD level of theory using different basis sets are collected in Tables~\ref{table-IV} and \ref{table-V}. These constants have been rounded off to three decimal places in these tables.\\

For comparison with the results reported in Ref.~\cite{Polet_2024}, obtained using the Fock-space coupled-cluster method with single and double excitations (FSCCSD), we employ a similar v2z basis set within a configuration space of $\pm$ \(2\)$E_h$. The computed constants for all molecules, along with the corresponding literature values, are presented in Table~\ref{table-IV}. For YbCu, our values of $W_d$ and $W_s$ computed using G-C1 are smaller by \(2.8\)\% compared to their values, while those obtained with G-C2 are larger by \(3.2\)\%. In the case of YbAg, the $W_d$ and $W_s$ constants exhibit differences of \(3\)\% (\(6.4\)\%) and \(1.4\)\% (\(5.5\)\%), respectively, for G-C1 (G-C2). These observations indicate that G-C1 provides comparable agreement with literature results for both YbCu and YbAg molecules. However, for YbAu, the constants obtained using G-C1 vary significantly. In contrast, G-C2 yields more reliable results for this system, with absolute differences of \(3.49\) $(\text{in}\,\, 10^{24} \,h\,\text{Hz}/e\,\text{cm})$ and \(0.67\) $(\text{in}\,\,h\,\text{kHz})$ for $W_d$ and $W_s$, respectively, compared to the FSCCSD results, even though the number of determinants in G-C2 is smaller than that in G-C1. The differences in the results of the two studies arise solely from distinct treatments of electron correlation effects. We have further examined the effect of the additional functions present in the cv2z basis set relative to v2z and found that they have a negligible impact on the computed constants. The final values of the interaction constants in Ref.~\cite{Polet_2024} have taken into account contributions from an expanded configuration space and triple excitations in the FSCC method. However, due to computational constraints, we have not been able to employ a comparable configuration space or include higher-order excitations in the present work.\\

We have discussed below the results obtained using larger basis sets (cv3z and cv4z) with distinct choices of the configuration space for the considered molecular systems, and are summarized in Table~\ref{table-V}. The specific selection of active electrons and virtual orbitals is made to achieve an optimal balance between computational cost and accuracy. We have observed that increasing the energy cutoff for limiting the virtual space results in only modest variations in the $\mathcal{P}$-odd \& $\mathcal{T}$-odd interaction constants for both YbCu and YbAg. For example, raising the cutoff from \(1\) $E_h$ to \(2\) $E_h$ in the cv3z basis for YbCu results in maximum changes of \(1.62\)\% (G-C1) and \(3.33\)\% (G-C2), whereas increasing the cutoff from \(1\) $E_h$ to \(1.5\) $E_h$ in the cv4z basis leads to deviations of \(1.57\)\% (G-C1) and \(2.81\)\% (G-C2). With identical energy cutoffs, the corresponding changes for YbAg are at most \(2.9\)\% (G-C1) and \(3.15\)\% (G-C2) for the cv3z basis, while \(2.26\)\% (G-C1) and \(2.62\)\% (G-C2) for the cv4z basis. The YbAu system, however, exhibits relatively larger variations upon increasing the virtual space cutoff, indicating a stronger sensitivity to the size of the virtual space compared to YbCu and YbAg.\\ 

The influence of the basis set size on the values of $W_d$ and $W_s$ has been assessed by comparing constants calculated with the configuration space fixed at $\pm$ \(1\) $E_h$ using the cv3z and cv4z basis sets. We have found that the maximum deviations in $W_d$ and $W_s$ for YbCu are \(0.52\)\% and \(0.15\)\%, respectively, considering either of the two GAS configurations (G-C1 or G-C2). For YbAg, the respective variations are less than \(1.06\)\% in $W_d$ and \(0.65\)\% in $W_s$. Notably, the change in the constants with increasing basis set size is substantial for YbAu. However, for the results obtained using the cv3z basis set with an expanded configuration space of \(-1\) $E_h$ to \(2\) $E_h$, the dependence on the basis set size is expected to be relatively small. Further, the results calculated using G-C1 differ from G-C2 values by a maximum of \(1.03\)\% for YbCu and \(1.87\)\% for YbAg. In particular, G-C2 results show better agreement with the available literature values for YbAu, as discussed earlier. Therefore, we recommend G-C2 results computed using the cv3z basis set, as highlighted in bold font in Table~\ref{table-V}.\\

Furthermore, the computed $\mathcal{P}$-odd \& $\mathcal{T}$-odd interaction constants have similar magnitudes for YbCu and YbAg, while they are much smaller for YbAu. A similar observation has been reported in Ref.~\cite{Polet_2024}. The comparatively small value for YbAu results from the counterbalancing of large, nearly equal contributions with opposite signs from the two constituent atoms. For YbCu and YbAg, the results fall within the range of values reported for the other polar molecules being investigated to probe $\mathcal{P}$-odd \& $\mathcal{T}$-odd effects~\cite{TFleig_2017,Gaul_2019,Gaul_2020}.

\subsection{Magnetic dipole hyperfine structure constants}
\begin{table*}
\caption{\label{table-VI}Magnitudes of the parallel ($A_\parallel$) and the perpendicular ($A_\perp$) components of the magnetic dipole HFS constants (\text{in}\,\,\text{MHz}) for the ground state of YbX molecules, calculated at the KRCISD level of theory using cv3z basis set and a configuration space spanning \(-1\) $E_h$ to \(2\) $E_h$. The results presented in bold fonts are our recommended values.} 
\begin{ruledtabular}
\begin{tabular}{cccccc}
&& \multicolumn{2}{c}{G-C1} & \multicolumn{2}{c}{G-C2} \\
\cline{3-4}\cline{5-6}
\multicolumn{6}{c}{\vspace{-3mm}}\\
Molecule & Atom & $A_\parallel$ & $A_\perp$ & $A_\parallel$ & $A_\perp$ \\
\hline
\multicolumn{6}{c}{\vspace{-3mm}}\\
\textbf{YbCu} & $^{171}$Yb & 3538.76 & 3185.82 & \textbf{3370.13} & \textbf{3079.54} \\
              & $^{173}$Yb & 974.94 & 877.70 & \textbf{928.48} & \textbf{848.42} \\
              & $^{63}$Cu & 1094.56 & 1093.45 & \textbf{1400.18} & \textbf{1402.22} \\
              & $^{65}$Cu & 1172.54 & 1171.35 & \textbf{1499.94} & \textbf{1502.11} \\
\hline
\multicolumn{6}{c}{\vspace{-3mm}}\\
\textbf{YbAg} & $^{171}$Yb & 3641.22 & 3312.99 & \textbf{3547.68} & \textbf{3273.27} \\
              & $^{173}$Yb & 1003.16 & 912.73 & \textbf{977.40} & \textbf{901.80} \\
              & $^{107}$Ag & 359.20 & 359.90 & \textbf{429.13} & \textbf{430.57} \\
              & $^{109}$Ag & 413.34 & 414.14 & \textbf{493.82} & \textbf{495.49} \\
\hline
\multicolumn{6}{c}{\vspace{-3mm}}\\
\textbf{YbAu} & $^{171}$Yb & 4683.30 & 4396.34 & \textbf{4852.62} & \textbf{4613.84} \\
              & $^{173}$Yb & 1290.26 & 1211.20 & \textbf{1336.91} & \textbf{1271.13} \\
              & $^{197}$Au & 273.71 & 270.92 & \textbf{304.15} & \textbf{300.95} \\
\end{tabular}
\end{ruledtabular}
\end{table*}

The results for the parallel ($A_\parallel$) and perpendicular ($A_\perp$) components of the magnetic dipole hyperfine structure constants (HFS) for different isotopes of the atoms constituting the molecules considered in this work are given in Table~\ref{table-VI}. We have employed the same basis set and configuration space for the HFS calculations as utilized for the recommended results of the $\mathcal{P}$-odd \& $\mathcal{T}$-odd interaction constants. To the best of our knowledge, these values are reported for the first time in our work. The accuracy of the present HFS results can only be assessed through future theoretical works or experimental measurements. This, in turn, will further validate the reliability of the computed $\mathcal{P}$-odd \& $\mathcal{T}$-odd constants, as the calculation of both requires a precise wavefunction in the vicinity of the nucleus.\\

We have calculated the HFS constants using both GAS configurations (G-C1 and G-C2), as shown in Table~\ref{table-VI}. However, to be consistent with the recommended interaction constants discussed above, we have carried out the analysis of the HFS constants computed using G-C2. We have observed that the absolute differences in the HFS components between $^{171}$Yb and $^{173}$Yb (\textit{viz.} $\Delta{A}_\parallel$, $\Delta{A}_\perp$), in MHz, are (\(2441.65\), \(2231.12\)), (\(2570.28\), \(2371.47\)), and (\(3515.71\), \(3342.71\)) in the molecular environments of YbCu, YbAg, and YbAu, respectively. For Cu and Ag isotopes in their corresponding molecular systems, the absolute differences in the components of HFS constants between the isotopes, in MHz, are (\(99.76\), \(99.89\)) and (\(64.69\), \(64.92\)), respectively.\\

The components of the HFS constants for the isotopes of Yb increase as we move from YbCu to YbAu. This indicates that the spin density near the Yb nucleus in YbCu is lower than that in YbAu. Further, we have analyzed the deviation of the HFS constants obtained using G-C1 from the final G-C2 results. The $A_\parallel$ and $A_\perp$ components of the HFS constants for Yb isotopes in the molecular environment of YbX molecules show a maximum variation of \(5\)\% between the two GAS configurations. For the isotopes of Cu, Ag, and Au, the corresponding deviations are at most \(22\)\%, \(16\)\%, and \(10\)\%, respectively. These differences reflect the impact of redistributing active orbitals among the GAS subspaces on the calculated HFS constants.\\

The knowledge of HFS is required for laser-cooling and atom-trapping experiments due to its influence on optical selection rules, and the transfer of momentum from photons to atoms~\cite{Phillips_1998}.\\

\section{\label{section5}Summary}

To summarize, we have carried out calculations of the symmetry violating interaction constants of YbX molecules using the relativistic KRCISD method. These computations are performed using the GAS technique with relativistic basis sets. The computed results are compared with the sole theoretical work available in the literature~\cite{Polet_2024}. We have found that the $W_d$ and $W_s$ constants are of similar magnitude for YbCu and YbAg, whereas YbAu exhibits smaller values. The increase in the size of the virtual orbital space as well as the basis set size for YbAu significantly affects the calculated values of the interaction constants. Further, we have investigated the impact of redistributing active orbitals among the GAS subspaces on the computed constants. This investigation reveals a maximum change in the $\mathcal{P}$-odd \& $\mathcal{T}$-odd constants of \(1.03\)\% for YbCu and \(1.87\)\% for YbAg, while a substantially larger change is observed for YbAu. To the best of our knowledge, no reported values of HFS constants hitherto exist for these molecules in the literature. We believe that the results presented in this work would be valuable for future theoretical and experimental investigations related to the search for the eEDM in these molecular systems.

\begin{acknowledgments}
We would like to thank the National Supercomputing Mission (NSM) for providing computing resources of ‘PARAM Ganga’ at the Indian Institute of Technology Roorkee, implemented by C-DAC and supported by the Ministry of Electronics and Information Technology (MeitY) and Department of Science and Technology (DST), Government of India. Some of the calculations reported in this work were performed on the high-performance computing facility available in the Department of Physics, IIT Roorkee, India. R.B. was supported by Polish National Science Centre Project No. 2021/41/B/ST2/00681. The research is also a part of the program of the National Laboratory FAMO in Toru\'n, Poland.
\end{acknowledgments}
\bibliography{PT}

@PREAMBLE{
 "\providecommand{\noopsort}[1]{}" 
 # "\providecommand{\singleletter}[1]{#1}%" 
}

@article{Polet_2024,
    author = {Polet, Johan David and Chamorro, Yuly and Pašteka, Lukáš F. and Hoekstra, Steven and Tomza, Michał and Borschevsky, Anastasia and Aucar, I. Agustín},
    title = {$\mathcal{P,T}$-odd effects in {YbCu}, {YbAg}, and {YbAu}},
    journal = {The Journal of Chemical Physics},
    volume = {161},
    number = {23},
    pages = {234302},
    year = {2024},
    month = {12},
    issn = {0021-9606},
    doi = {10.1063/5.0235522},
    url = {https://doi.org/10.1063/5.0235522}
}

@article{Yuan_2024,
  title = {Forming ultracold {YbAg} molecules via photoassociation predicted from \textit{ab initio} calculations},
  author = {Yuan, Xiang and Liu, Yong},
  journal = {Phys. Rev. A},
  volume = {110},
  issue = {6},
  pages = {062813},
  numpages = {10},
  year = {2024},
  month = {Dec},
  publisher = {American Physical Society},
  doi = {10.1103/PhysRevA.110.062813},
  url = {https://link.aps.org/doi/10.1103/PhysRevA.110.062813}
}

@article{Bala_2023,
  title = { Effective electric field associated with the electric dipole moment of the electron for {TlF}$^+$},
  author = {Bala, R. and Prasannaa, V. S. and Abe, M. and Das, B. P.},
  journal = {The European Physical Journal Plus},
  volume = {138},
  issue = {5},
  pages = {478},
  numpages = {10},
  year = {2023},
  doi = {10.1140/epjp/s13360-023-04115-w}
}

@article{Bala_2020,
doi = {10.1088/1361-6455/ab87e8},
url = {https://doi.org/10.1088/1361-6455/ab87e8},
year = {2020},
month = {jun},
publisher = {IOP Publishing},
volume = {53},
number = {13},
pages = {135101},
author = {Bala, Renu and Nataraj, H S and Nayak, Malaya K},
title = {Calculations of $\mathcal{P}$ and $\mathcal{T}$-odd interaction constants of alkaline-earth monofluorides using {KRCI} method},
journal = {Journal of Physics B: Atomic, Molecular and Optical Physics}
}

@book{Szabo,
title={Modern Quantum Chemistry, Introduction to advanced electronic structure theory},
author={Szabo, A. and Ostlund N. S.},
year={1996},
publisher={Dover Publications, Inc. Mineola, New York}
}

@article{Fleig_2014,
title = {Electron electric dipole moment and hyperfine interaction constants for {ThO}},
journal = {Journal of Molecular Spectroscopy},
volume = {300},
pages = {16-21},
year = {2014},
issn = {0022-2852},
doi = {https://doi.org/10.1016/j.jms.2014.03.017},
url = {https://www.sciencedirect.com/science/article/pii/S0022285214000782},
author = {Timo Fleig and Malaya K. Nayak}
}

@article{Fleig_2013,
  title = {Electron electric-dipole-moment interaction constant for {HfF}${}^{+}$ from relativistic correlated all-electron theory},
  author = {Fleig, Timo and Nayak, Malaya K.},
  journal = {Phys. Rev. A},
  volume = {88},
  issue = {3},
  pages = {032514},
  numpages = {6},
  year = {2013},
  month = {Sep},
  publisher = {American Physical Society},
  doi = {10.1103/PhysRevA.88.032514},
  url = {https://link.aps.org/doi/10.1103/PhysRevA.88.032514}
}

@article{Prasannaa_2015,
  title = {Mercury Monohalides: Suitability for Electron Electric Dipole Moment Searches},
  author = {Prasannaa, V. S. and Vutha, A. C. and Abe, M. and Das, B. P.},
  journal = {Phys. Rev. Lett.},
  volume = {114},
  issue = {18},
  pages = {183001},
  numpages = {6},
  year = {2015},
  month = {May},
  publisher = {American Physical Society},
  doi = {10.1103/PhysRevLett.114.183001}
}

@article{Sunaga_2019,
  title = {Ultracold mercury--alkali-metal molecules for electron-electric-dipole-moment searches},
  author = {Sunaga, A. and Prasannaa, V. S. and Abe, M. and Hada, M. and Das, B. P.},
  journal = {Phys. Rev. A},
  volume = {99},
  issue = {4},
  pages = {040501},
  numpages = {6},
  year = {2019},
  month = {Apr},
  publisher = {American Physical Society},
  doi = {10.1103/PhysRevA.99.040501}
}

@article{Nayak_2007,
  title = {\textit{Ab initio} calculation of the electron-nucleus scalar-pseudoscalar interaction constant ${W}_{s}$ in heavy polar molecules},
  author = {Nayak, Malaya K. and Chaudhuri, Rajat K. and Das, B. P.},
  journal = {Phys. Rev. A},
  volume = {75},
  issue = {2},
  pages = {022510},
  numpages = {3},
  year = {2007},
  month = {Feb},
  publisher = {American Physical Society},
  doi = {10.1103/PhysRevA.75.022510}
}

@article{Chamorro_2024,
  title = {Enhanced parity and time-reversal-symmetry violation in diatomic molecules: {LaO}, {LaS}, and {LuO}},
  author = {Chamorro, Yuly and Flambaum, Victor V. and Garcia Ruiz, Ronald F. and Borschevsky, Anastasia and Pa{\v{s}}teka, Luk{\'a}{\v{s}} F.},
  journal = {Phys. Rev. A},
  volume = {110},
  issue = {4},
  pages = {042806},
  numpages = {12},
  year = {2024},
  month = {Oct},
  publisher = {American Physical Society},
  doi = {10.1103/PhysRevA.110.042806}
}

@article{Denis_2015,
doi = {10.1088/1367-2630/17/4/043005},
url = {https://doi.org/10.1088/1367-2630/17/4/043005},
year = {2015},
month = {apr},
publisher = {IOP Publishing},
volume = {17},
number = {4},
pages = {043005},
author = {Denis, Malika and Nørby, Morten S and Jensen, Hans Jørgen Aa and Gomes, André Severo Pereira and Nayak, Malaya K and Knecht, Stefan and Fleig, Timo},
title = {Theoretical study on {ThF}$^{+}$, a prospective system in search of time-reversal violation},
journal = {New Journal of Physics}
}

@book{Lindgren_2012,
  title={Atomic many-body theory},
  author={Lindgren, Ingvar and Morrison, John},
  volume={3},
  year={2012},
  publisher={Springer Science \& Business Media}
}

@article{Sasmal_2015,
  title = {Relativistic extended-coupled-cluster method for the magnetic hyperfine structure constant},
  author = {Sasmal, Sudip and Pathak, Himadri and Nayak, Malaya K. and Vaval, Nayana and Pal, Sourav},
  journal = {Phys. Rev. A},
  volume = {91},
  issue = {2},
  pages = {022512},
  numpages = {10},
  year = {2015},
  month = {Feb},
  publisher = {American Physical Society},
  doi = {10.1103/PhysRevA.91.022512}
}

@article{Tulukdar_2019,
    author = {Talukdar, Kaushik and Nayak, Malaya K. and Vaval, Nayana and Pal, Sourav},
    title = {Relativistic coupled-cluster investigation of parity ($\mathcal{P}$) and time-reversal ($\mathcal{T}$) symmetry violations in {HgF}},
    journal = {The Journal of Chemical Physics},
    volume = {150},
    number = {8},
    pages = {084304},
    year = {2019},
    month = {02},
    issn = {0021-9606},
    doi = {10.1063/1.5083000},
    url = {https://doi.org/10.1063/1.5083000}
}

@article{Sasmal_2016,
  author  = {Sasmal, Sudip and Talukdar, Kaushik and Nayak, Malaya K. and Vaval, Nayana and Pal, Sourav},
  title   = {Calculation of hyperfine structure constants of small molecules using {Z}-vector method in the relativistic coupled-cluster framework},
  journal = {Journal of Chemical Sciences},
  year    = {2016},
  volume  = {128},
  number  = {10},
  pages   = {1671--1675},
  doi     = {10.1007/s12039-016-1174-1}
}

@article{Knecht_2010,
    author = {Knecht, Stefan and Jensen, Hans Jørgen Aa. and Fleig, Timo},
    title = {Large-scale parallel configuration interaction. {II}. {T}wo- and four-component double-group general active space implementation with application to {BiH}},
    journal = {The Journal of Chemical Physics},
    volume = {132},
    number = {1},
    pages = {014108},
    year = {2010},
    month = {01},
    issn = {0021-9606},
    doi = {10.1063/1.3276157},
    url = {https://doi.org/10.1063/1.3276157}
}

@article{Fleig_2016,
  title = {{TaN}, a molecular system for probing $\mathcal{P},\mathcal{T}$-violating hadron physics},
  author = {Fleig, Timo and Nayak, Malaya K. and Kozlov, Mikhail G.},
  journal = {Phys. Rev. A},
  volume = {93},
  issue = {1},
  pages = {012505},
  numpages = {10},
  year = {2016},
  month = {Jan},
  publisher = {American Physical Society},
  doi = {10.1103/PhysRevA.93.012505},
  url = {https://link.aps.org/doi/10.1103/PhysRevA.93.012505}
}

@article{DIRAC,
      Title = {\textit {{DIRAC}, a relativistic ab initio electronic structure program, Release {DIRAC}23}},
      year= {2023 (http://www.diracprogram.org)},
      author= {R. Bast and A. S. P. Gomes and T. Saue and L. Visscher and H. J. Aa. Jensen and with contributions from I. A. Aucar and V. Bakken and C. Chibueze and J. Creutzberg and K. G. Dyall and S. Dubillard and U. Ekström and E. Eliav and T. Enevoldsen and E. Faßhauer and T. Fleig and O. Fossgaard and L. Halbert and E. D. Hedegård and T. Helgaker and B. Helmich-Paris and J. Henriksson and M. van Horn and M. Iliaš and Ch. R. Jacob and S. Knecht and S. Komorovský and O. Kullie and J. K. Lærdahl and C. V. Larsen and Y. S. Lee and N. H. List and H. S. Nataraj and M. K. Nayak and P. Norman and A. Nyvang and G. Olejniczak and J. Olsen and J. M. H. Olsen and A. Papadopoulos and Y. C. Park and J. K. Pedersen and M. Pernpointner and J. V. Pototschnig and R. Di Remigio and M. Repisky and K. Ruud and P. Sałek and B. Schimmelpfennig and B. Senjean and A. Shee and J. Sikkema and A. Sunaga and A. J. Thorvaldsen and J. Thyssen and J. van Stralen and M. L. Vidal and S. Villaume and O. Visser and T. Winther and S. Yamamoto and X. Yuan},
      journal= {},
     volume = {},
      pages = {},
      url= {http://www.diracprogram.org},
      note={available at \url{https://doi.org/10.5281/zenodo.7670749}}
}

@article{TFleig_2006,
    author = {Fleig, Timo and Jensen, Hans Jørgen Aa. and Olsen, Jeppe and Visscher, Lucas},
    title = {The generalized active space concept for the relativistic treatment of electron correlation. {III}. {L}arge-scale configuration interaction and multiconfiguration self-consistent-field four-component methods with application to {UO}$_{2}$},
    journal = {The Journal of Chemical Physics},
    volume = {124},
    number = {10},
    pages = {104106},
    year = {2006},
    month = {03},
    issn = {0021-9606},
    doi = {10.1063/1.2176609},
    url = {https://doi.org/10.1063/1.2176609}
}

@article{Dyall_2007,
  author  = {Dyall, Kenneth G.},
  title   = {Relativistic double-zeta, triple-zeta, and quadruple-zeta basis sets for the 4d elements {Y}--{Cd}},
  journal = {Theoretical Chemistry Accounts},
  year    = {2007},
  volume  = {117},
  number  = {4},
  pages   = {483--489},
  doi     = {10.1007/s00214-006-0174-5},
  issn    = {1432-2234}
}

@article{Gomes_2010,
  author  = {Gomes, Andr{\'e} S. P. and Dyall, Kenneth G. and Visscher, Lucas},
  title   = {Relativistic double-zeta, triple-zeta, and quadruple-zeta basis sets for the lanthanides {La}--{Lu}},
  journal = {Theoretical Chemistry Accounts},
  year    = {2010},
  volume  = {127},
  number  = {4},
  pages   = {369--381},
  doi     = {10.1007/s00214-009-0725-7},
  issn    = {1432-2234}
}

@article{Dyall_2010,
  author  = {Dyall, Kenneth G. and Gomes, Andre S. P.},
  title   = {Revised relativistic basis sets for the 5d elements {Hf}--{Hg}},
  journal = {Theoretical Chemistry Accounts},
  year    = {2010},
  volume  = {125},
  number  = {1},
  pages   = {97--100},
  doi     = {10.1007/s00214-009-0717-7},
  issn    = {1432-2234}
}

@article{Dyall_2023,
author = {Dyall, Kenneth G. and Tecmer, Paweł and Sunaga, Ayaki},
title = {Diffuse Basis Functions for Relativistic s and d Block Gaussian Basis Sets},
journal = {Journal of Chemical Theory and Computation},
volume = {19},
number = {1},
pages = {198-210},
year = {2023},
doi = {10.1021/acs.jctc.2c01050},
URL = {https://doi.org/10.1021/acs.jctc.2c01050}
}

@misc{isotopes,
  howpublished = {\url{https://www.webelements.com/isotopes.html}}
}

@article{Ginges_2004,
title = {Violations of fundamental symmetries in atoms and tests of unification theories of elementary particles},
journal = {Physics Reports},
volume = {397},
number = {2},
pages = {63-154},
year = {2004},
issn = {0370-1573},
doi = {https://doi.org/10.1016/j.physrep.2004.03.005},
url = {https://www.sciencedirect.com/science/article/pii/S0370157304001322},
author = {J.S.M. Ginges and V.V. Flambaum}
}

@article{Safronova_2018,
  title = {Search for new physics with atoms and molecules},
  author = {Safronova, M. S. and Budker, D. and DeMille, D. and Kimball, Derek F. Jackson and Derevianko, A. and Clark, Charles W.},
  journal = {Rev. Mod. Phys.},
  volume = {90},
  issue = {2},
  pages = {025008},
  numpages = {106},
  year = {2018},
  month = {Jun},
  publisher = {American Physical Society},
  doi = {10.1103/RevModPhys.90.025008},
  url = {https://link.aps.org/doi/10.1103/RevModPhys.90.025008}
}

@article{Fukuyama_2012,
author = {Fukuyama, TAKESHI},
title = {SEARCHING FOR NEW PHYSICS BEYOND THE STANDARD MODEL IN ELECTRIC DIPOLE MOMENT},
journal = {International Journal of Modern Physics A},
volume = {27},
number = {16},
pages = {1230015},
year = {2012},
doi = {10.1142/S0217751X12300153},
URL = {https://doi.org/10.1142/S0217751X12300153}
}

@article{Landau_1957,
title = {On the conservation laws for weak interactions},
journal = {Nuclear Physics},
volume = {3},
number = {1},
pages = {127-131},
year = {1957},
issn = {0029-5582},
doi = {https://doi.org/10.1016/0029-5582(57)90061-5},
url = {https://www.sciencedirect.com/science/article/pii/0029558257900615},
author = {L. Landau}
}

@article{Purcell_1950,
  title = {On the Possibility of Electric Dipole Moments for Elementary Particles and Nuclei},
  author = {Purcell, E. M. and Ramsey, N. F.},
  journal = {Phys. Rev.},
  volume = {78},
  issue = {6},
  pages = {807--807},
  numpages = {0},
  year = {1950},
  month = {Jun},
  publisher = {American Physical Society},
  doi = {10.1103/PhysRev.78.807},
  url = {https://link.aps.org/doi/10.1103/PhysRev.78.807}
}

@article{Luders_2000,
title = {Proof of the {TCP} Theorem},
journal = {Annals of Physics},
volume = {281},
number = {1},
pages = {1004-1018},
year = {2000},
issn = {0003-4916},
doi = {https://doi.org/10.1006/aphy.2000.6027},
url = {https://www.sciencedirect.com/science/article/pii/S0003491600960275},
author = {Gerhart Lüders}
}

@article{Kazarian_1992,
title = {Cosmological lower bound on the {EDM} of the electron},
journal = {Physics Letters B},
volume = {276},
number = {1},
pages = {131-134},
year = {1992},
issn = {0370-2693},
doi = {https://doi.org/10.1016/0370-2693(92)90552-F},
url = {https://www.sciencedirect.com/science/article/pii/037026939290552F},
author = {A.M. Kazarian and S.V. Kuzmin and M.E. Shaposhnikov}
}

@article{Virdee_2016,
    author = {Virdee, T. S.},
    title = {Beyond the standard model of particle physics},
    journal = {Philosophical Transactions of the Royal Society A: Mathematical, Physical and Engineering Sciences},
    volume = {374},
    number = {2075},
    pages = {20150259},
    year = {2016},
    month = {08},
    doi = {10.1098/rsta.2015.0259},
    url = {https://doi.org/10.1098/rsta.2015.0259}
}

@article{Chupp_2019,
  title = {Electric dipole moments of atoms, molecules, nuclei, and particles},
  author = {Chupp, T. E. and Fierlinger, P. and Ramsey-Musolf, M. J. and Singh, J. T.},
  journal = {Rev. Mod. Phys.},
  volume = {91},
  issue = {1},
  pages = {015001},
  numpages = {55},
  year = {2019},
  month = {Jan},
  publisher = {American Physical Society},
  doi = {10.1103/RevModPhys.91.015001},
  url = {https://link.aps.org/doi/10.1103/RevModPhys.91.015001}
}

@article{Quiney_1998,
doi = {10.1088/0953-4075/31/3/003},
url = {https://doi.org/10.1088/0953-4075/31/3/003},
year = {1998},
month = {feb},
publisher = {},
volume = {31},
number = {3},
pages = {L85},
author = {H M Quiney and H Skaane and I P Grant},
title = {Hyperfine and  \textit{PT}-odd effects in {YbF}},
journal = {Journal of Physics B: Atomic, Molecular and Optical Physics}
}

@article{Sandars_1967,
  title = {Measurability of the Proton Electric Dipole Moment},
  author = {Sandars, P. G. H.},
  journal = {Phys. Rev. Lett.},
  volume = {19},
  issue = {24},
  pages = {1396--1398},
  numpages = {0},
  year = {1967},
  month = {Dec},
  publisher = {American Physical Society},
  doi = {10.1103/PhysRevLett.19.1396},
  url = {https://link.aps.org/doi/10.1103/PhysRevLett.19.1396}
}

@article{Andreev_2018,
  author       = {Andreev, V. and Ang, D. G. and DeMille, D. and Doyle, J. M.
                  and Gabrielse, G. and Haefner, J. and Hutzler, N. R.
                  and Lasner, Z. and Meisenhelder, C. and O'Leary, B. R.
                  and Panda, C. D. and West, A. D. and West, E. P.
                  and Wu, X. and {ACME Collaboration}},
  title        = {Improved limit on the electric dipole moment of the electron},
  journal      = {Nature},
  volume       = {562},
  number       = {7727},
  pages        = {355--360},
  year         = {2018},
  doi          = {10.1038/s41586-018-0599-8},
  url          = {https://doi.org/10.1038/s41586-018-0599-8},
  issn         = {1476-4687}
}

@article{Roussy_2023,
author = {Tanya S. Roussy  and Luke Caldwell  and Trevor Wright  and William B. Cairncross  and Yuval Shagam  and Kia Boon Ng  and Noah Schlossberger  and Sun Yool Park  and Anzhou Wang  and Jun Ye  and Eric A. Cornell },
title = {An improved bound on the electron’s electric dipole moment},
journal = {Science},
volume = {381},
number = {6653},
pages = {46-50},
year = {2023},
doi = {10.1126/science.adg4084},
URL = {https://www.science.org/doi/abs/10.1126/science.adg4084}
}

@article{Fitch_2020,
doi = {10.1088/2058-9565/abc931},
url = {https://doi.org/10.1088/2058-9565/abc931},
year = {2020},
month = {dec},
publisher = {IOP Publishing},
volume = {6},
number = {1},
pages = {014006},
author = {Fitch, N J and Lim, J and Hinds, E A and Sauer, B E and Tarbutt, M R},
title = {Methods for measuring the electron’s electric dipole moment using ultracold {YbF} molecules},
journal = {Quantum Science and Technology}
}

@article{Aggarwal_2018,
  author       = {Aggarwal, Parul and Bethlem, Hendrick L. and Borschevsky, Anastasia
                  and Denis, Malika and Esajas, Kevin and Haase, Pi A. B.
                  and Hao, Yongliang and Hoekstra, Steven and Jungmann, Klaus
                  and Meijknecht, Thomas B. and Mooij, Maarten C.
                  and Timmermans, Rob G. E. and Ubachs, Wim
                  and Willmann, Lorenz and Zapara, Artem
                  and {NL-eEDM Collaboration}},
  title        = {Measuring the electric dipole moment of the electron in {BaF}},
  journal      = {The European Physical Journal D},
  volume       = {72},
  number       = {11},
  pages        = {197},
  year         = {2018},
  doi          = {10.1140/epjd/e2018-90192-9},
  url          = {https://doi.org/10.1140/epjd/e2018-90192-9},
  issn         = {1434-6079}
}

@Article{Ramanuj_2021,
AUTHOR = {Mitra, Ramanuj and Prasannaa, V. Srinivasa and Sahoo, Bijaya K. and Hutzler, Nicholas R. and Abe, Minori and Das, Bhanu Pratap},
TITLE = {Study of {HgOH} to Assess Its Suitability for Electron Electric Dipole Moment Searches},
JOURNAL = {Atoms},
VOLUME = {9},
YEAR = {2021},
NUMBER = {1},
ARTICLE-NUMBER = {7},
URL = {https://www.mdpi.com/2218-2004/9/1/7},
ISSN = {2218-2004}
}

@article{Fleig_2021,
doi = {10.1088/1367-2630/ac3619},
url = {https://doi.org/10.1088/1367-2630/ac3619},
year = {2021},
month = {nov},
publisher = {IOP Publishing},
volume = {23},
number = {11},
pages = {113039},
author = {Fleig, Timo and DeMille, David},
title = {Theoretical aspects of radium-containing molecules amenable to assembly from laser-cooled atoms for new physics searches},
journal = {New Journal of Physics}
}

@article{Kudashov_2014,
  title = {\textit{Ab initio} study of radium monofluoride ({RaF}) as a candidate to search for parity- and time-and-parity--violation effects},
  author = {Kudashov, A. D. and Petrov, A. N. and Skripnikov, L. V. and Mosyagin, N. S. and Isaev, T. A. and Berger, R. and Titov, A. V.},
  journal = {Phys. Rev. A},
  volume = {90},
  issue = {5},
  pages = {052513},
  numpages = {5},
  year = {2014},
  month = {Nov},
  publisher = {American Physical Society},
  doi = {10.1103/PhysRevA.90.052513},
  url = {https://link.aps.org/doi/10.1103/PhysRevA.90.052513}
}

@article{Fazil_2019,
  title = {{RaH} as a potential candidate for electron electric-dipole-moment searches},
  author = {Fazil, N. M. and Prasannaa, V. S. and Latha, K. V. P. and Abe, M. and Das, B. P.},
  journal = {Phys. Rev. A},
  volume = {99},
  issue = {5},
  pages = {052502},
  numpages = {5},
  year = {2019},
  month = {May},
  publisher = {American Physical Society},
  doi = {10.1103/PhysRevA.99.052502},
  url = {https://link.aps.org/doi/10.1103/PhysRevA.99.052502}
}

@article{Verma_2020,
  title = {Electron Electric Dipole Moment Searches Using Clock Transitions in Ultracold Molecules},
  author = {Verma, Mohit and Jayich, Andrew M. and Vutha, Amar C.},
  journal = {Phys. Rev. Lett.},
  volume = {125},
  issue = {15},
  pages = {153201},
  numpages = {5},
  year = {2020},
  month = {Oct},
  publisher = {American Physical Society},
  doi = {10.1103/PhysRevLett.125.153201},
  url = {https://link.aps.org/doi/10.1103/PhysRevLett.125.153201}
}

@article{Tomza_2021,
doi = {10.1088/1367-2630/ac1696},
url = {https://doi.org/10.1088/1367-2630/ac1696},
year = {2021},
month = {aug},
publisher = {IOP Publishing},
volume = {23},
number = {8},
pages = {085003},
author = {Tomza, Michał},
title = {Interaction potentials, electric moments, polarizabilities, and chemical reactions of {YbCu}, {YbAg}, and {YbAu} molecules},
journal = {New Journal of Physics}
}

@article{Gaul_2019,
  title = {Systematic study of relativistic and chemical enhancements of $\mathcal{P},\mathcal{T}$-odd effects in polar diatomic radicals},
  author = {Gaul, Konstantin and Marquardt, Sebastian and Isaev, Timur and Berger, Robert},
  journal = {Phys. Rev. A},
  volume = {99},
  issue = {3},
  pages = {032509},
  numpages = {17},
  year = {2019},
  month = {Mar},
  publisher = {American Physical Society},
  doi = {10.1103/PhysRevA.99.032509},
  url = {https://link.aps.org/doi/10.1103/PhysRevA.99.032509}
}

@article{TFleig_2017,
  title = {$\mathcal{P},\mathcal{T}$-odd and magnetic hyperfine-interaction constants and excited-state lifetime for {HfF}${}^{+}$},
  author = {Fleig, Timo},
  journal = {Phys. Rev. A},
  volume = {96},
  issue = {4},
  pages = {040502},
  numpages = {4},
  year = {2017},
  month = {Oct},
  publisher = {American Physical Society},
  doi = {10.1103/PhysRevA.96.040502},
  url = {https://link.aps.org/doi/10.1103/PhysRevA.96.040502}
}

@article{Gaul_2020,
  title = {\textit{Ab initio} study of parity and time-reversal violation in laser-coolable triatomic molecules},
  author = {Gaul, Konstantin and Berger, Robert},
  journal = {Phys. Rev. A},
  volume = {101},
  issue = {1},
  pages = {012508},
  numpages = {6},
  year = {2020},
  month = {Jan},
  publisher = {American Physical Society},
  doi = {10.1103/PhysRevA.101.012508},
  url = {https://link.aps.org/doi/10.1103/PhysRevA.101.012508}
}

@article{Phillips_1998,
  title = {Nobel {L}ecture: Laser cooling and trapping of neutral atoms},
  author = {Phillips, William D.},
  journal = {Rev. Mod. Phys.},
  volume = {70},
  issue = {3},
  pages = {721--741},
  numpages = {0},
  year = {1998},
  month = {Jul},
  publisher = {American Physical Society},
  doi = {10.1103/RevModPhys.70.721},
  url = {https://link.aps.org/doi/10.1103/RevModPhys.70.721}
}

@article{Skripnikov_2013,
    author = {Skripnikov, L. V. and Petrov, A. N. and Titov, A. V.},
    title = {Communication: Theoretical study of {ThO} for the electron electric dipole moment search},
    journal = {The Journal of Chemical Physics},
    volume = {139},
    number = {22},
    pages = {221103},
    year = {2013},
    month = {12},
    issn = {0021-9606},
    doi = {10.1063/1.4843955},
    url = {https://doi.org/10.1063/1.4843955}
}

@article{Abe_2014,
  title = {Application of relativistic coupled-cluster theory to the effective electric field in {YbF}},
  author = {Abe, M. and Gopakumar, G. and Hada, M. and Das, B. P. and Tatewaki, H. and Mukherjee, D.},
  journal = {Phys. Rev. A},
  volume = {90},
  issue = {2},
  pages = {022501},
  numpages = {5},
  year = {2014},
  month = {Aug},
  publisher = {American Physical Society},
  doi = {10.1103/PhysRevA.90.022501},
  url = {https://link.aps.org/doi/10.1103/PhysRevA.90.022501}
}
\end{document}